\documentstyle[11pt,paspconf,epsf]{article}

\begin{document}

\title{BaSeL: a library of synthetic spectra and colours for GAIA} 

\author{E. Lastennet}
\affil{Depto. de Astronomia, UFRJ, Rio de Janeiro, Brazil}

\author{T. Lejeune}
\affil{Observat\'orio Astron\'omico, Coimbra, Portugal}

\author{E. Oblak}
\affil{Observatoire de Besan\c{c}on, France} 

\author{P. Westera, R. Buser}
\affil{Astronomisches Institut der Universit\"at Basel,
Switzerland }

\begin{abstract}
The BaseL Stellar Library (BaSeL) is a library of synthetic spectra
which has already been used in various astrophysical applications
(stellar clusters studies, characterization and choice of the COROT
potential targets, eclipsing binaries,
...).
This library could provide useful indications to 1) choose the best photometric
system for the GAIA strategy by evaluating their expected performances and
2) apply systematically the BaSeL models for any sample of GAIA targets.
In this context, we describe one of the future developments
of the BaSeL interactive web site to probe the GAIA photometric data:
an automatic determination of atmospheric parameters from observed colours.

\end{abstract}


\section{Brief description of the BaSeL model and its interactive server}
The Basel Stellar Library (BaSeL) is a library of theoretical spectra
corrected to provide synthetic colours consistent with empirical colour-temperature
calibrations at all wavelengths from the near-UV to the far-IR (see
Lejeune et al. 1997, 1998 for a complete description,
and Westera et al. 1999 for the most recent version).
These model spectra cover a large range of fundamental parameters
(2000 $\leq$ T$_{\rm eff}$ $\leq$ 50,000 K, $-$5 $\leq$ [Fe/H] $\leq$ 1
and $-$1.02 $\leq$ log g $\leq$ 5.5) and hence
allow to investigate a large panel of multi-wavelength astrophysical
questions, as briefly reviewed in the next section.
Since they are based on synthetic spectra, they can in principle be used in
many photometric systems taken either individually or collectively,
and this is another major advantage of these models.
The "BaSeL interactive server" is the web version of the BaSeL models
({\tt http://www.astro.mat.uc.pt/BaSeL/}).
This server is under development and
the photometric systems presently available in interactive mode
are: Geneva, Washington, Johnson-Cousins, Str\"omgren, HST-WFPC2,
photographic RGU, and EROS.
All details about this server will be given elsewhere.


\section{Astrophysical applications}
The BaSeL models are currently used in an increasing number of
astrophysical studies.
It is far beyond the scope of this work to cite all the studies
but some examples of application cover 
globular clusters (e.g. Bruzual et al. 1997, Weiss \& Salaris 1999, 
Kurth et al. 1999, Lejeune \& Buser 1999), 
open clusters (e.g. Lastennet 1998, Pols et al. 1998),
preparation of the COROT space mission (Lastennet et al. 2001a),
HR diagram of the Hyades (e.g. Lastennet et al. 1999b,
Lebreton et al. 2001),
blue stragglers (e.g. Deng et al. 1999),
AGB stars (e.g. Lastennet et al. 2001b),
eclipsing binary stars (e.g. Lastennet et al. 1999a),
stars in the Small Magellanic Cloud (Cordier et al. 2000),
and clusters of galaxies (e.g. Steindling et al. 2001). 
Moreover, the photometric systems available in BaSeL  
can be used independently 
(e.g. the photographic RGU system, Buser et al. 2000)
or collectively 
(e.g. Johnson {\it and} Str\"omgren, Lastennet et al. 2001a). 
Finally, 
the BaSeL library has been used recently to
evaluate the performance of the proposed GAIA photometric
system 3G (Sudzius et al. 2001).

\section{The GAIA mission}
The GAIA mission, with a launch date in 2010-2012, is an approved Cornerstone
mission by ESA  designed to solve many of the most
fundamental challenges in
astronomy: to determine the composition,
formation and evolution of our Galaxy.
The core science case for GAIA requires measurement of luminosity, effective
temperature, mass, age, and composition for the stellar populations in our
own Galaxy and in its nearest galaxy neighbours. These quantities
can be derived from the spectral energy distribution of the stars, through
multi-band photometry.
Because the GAIA photometric system must be able to classify stars
across the entire HR diagram, as well as to identify peculiar
objects, none of the existing photometric systems satisfy all the
GAIA requirements, and a new system has to be defined
(standard wavebands and sets of standard stars).
Moreover, a recent study from the Vilnius GAIA group
(Vansevicius et al. 2001),
comparing the performance of the proposed GAIA photometric systems 1F,
2A \& 3G, seems to indicate that no optimal photometric system for GAIA
is proposed to date.
Covering a large spectral domain, extending from the UV to the far-IR,
BaSeL is adapted to perform simulations with the (new) proposed
photometric systems,
and should help to choose the most efficient one.

\section{One of the future developments of the BaSeL web site for GAIA}
In its actual version, the BaSeL models can already provide
relevant information to select between the proposed GAIA photometric
systems.
Given a set of effective temperature, metallicity and surface gravity
(T$_{\rm eff}$, [Fe/H], log g), the BaSeL models provide colours that
can be directly compared with photometric observations of stellar populations.
Alternatively, the inverse method (Lastennet et al. 1999a, LLWB99) would be
very useful to derive the atmospheric parameters {\it from} the observed
colours.
LLWB99
applied this method to derive simultaneously the
T$_{\rm eff}$ and metallicity of a sample of
eclipsing  binary stars (EBs) with photometric data in the Str\"omgren system.
They obtained very good results from the BaSeL models, in agreement with
the T$_{\rm eff}$ derived from accurate HIPPARCOS data
and with spectroscopic determination of metallicity (when available).
However, while they fix log g
because this quantity is
known with an excellent accuracy for their sample of EBs, we intend to use a
more general $\chi^2$-minimization algorithm for the GAIA targets in order to
derive the 3 atmospheric parameters:
\begin{equation}
 \chi^2 (T_{\rm eff}, [Fe/H], log\, g)  = \sum_{i=1}^{n} \left[
\left(\frac{\rm  GPI(i)_{\rm   syn} -  GPI(i)}{\sigma(\rm
GPI(i))}\right)^2  \right],
\end{equation}
where $n$ is the number of photometric indices,  
GPI(i) is a selected
{\bf G}AIA {\bf P}hoto\-metric {\bf I}ndex and GPI(i)$_{\rm syn}$ is the BaSeL
synthetic  index in the GAIA photometric system.
The uncertainty on the index measurements - $\sigma$(\rm GPI(i)) -
will provide confidence contours on the (T$_{\rm eff}$, [Fe/H], log g)
results. 
The interstellar absorption can either be included in equation 
(1) as an additional parameter (see an example in LLWB99, sect. 3.3.2) 
or can be determined by independent methods (see e.g. Lastennet et al. 2001a).
\begin{figure}
\plotfiddle{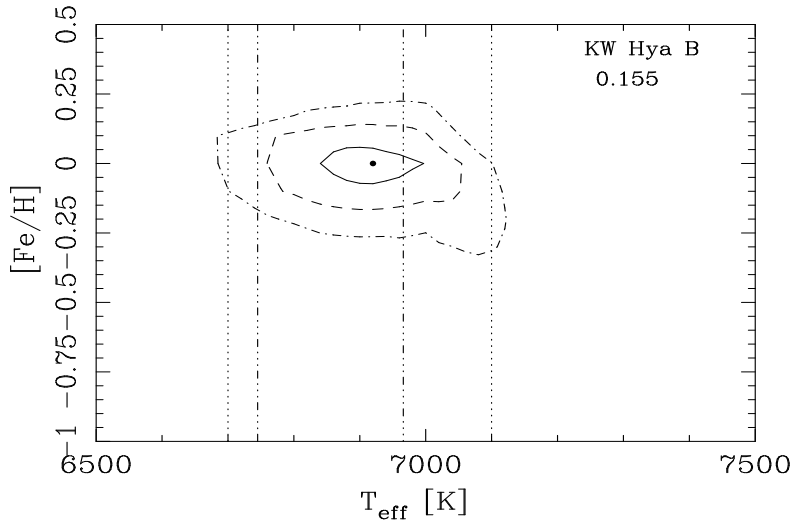}{6.cm}{-90}{40}{35}{-170}{200}
 \caption{
 \small
 Simultaneous (T$_{\rm eff}$, [Fe/H]) determination for the star
 KW Hya B (secondary component of an EB)
 from Str\"omgren photometry. Confidence regions (1, 2 and 3-$\sigma$)
 are derived from the synthetic BaSeL photometry.
 An estimation of the quality of the best fit ($\chi^2$-value) is quoted
 in the right upper corner.
 The T$_{\rm eff}$ from Andersen (1991)
(vertical dotted lines) and Ribas et al. (2000)
(vertical dot-dashed lines)
 are also shown for comparison.
\normalsize
 }
\end{figure}
An obvious advantage of this algorithm is to know the quality of the fit: 
an example is given in a T$_{\rm eff}$-[Fe/H] diagram (Fig. 1) 
where comparisons with previous T$_{\rm eff}$ determinations show
a good agreement. Note that the BaSeL models also provide an [Fe/H]
determination in Fig. 1.
Other examples, including bad fits, can be found in
Lastennet et al. (1999a, 2001a).
In the context of the GAIA mission, we propose to develop an
automatic tool based on this extended method of Lastennet et al. to complete
the facilities of the {\it BaSeL server}.
Provided that the new GAIA pass-bands (e.g. for the 1F system: 4 broad
and 11 intermediate bands covering the spectral range 280 to 920 nm)
are implemented in the BaSeL models, this new tool will provide automatically
(T$_{\rm eff}$, log g, [Fe/H]) estimates and uncertainties for the stars
observed with the GAIA photometric bands.
An automatic and efficient treatment of any sample of GAIA targets will
be possible.

\section{Conclusion}
Covering a large spectral domain, extending from the UV to the far-infrared,
the BaSeL models are adapted to perform simulations with the proposed
GAIA photometric systems and should help to choose the most efficient one.
We present and discuss a proposition to develop an automatic method,
already used with success for COROT potential targets (Lastennet et
al. 2001a), for a systematic determination of fundamental parameters
from BaSeL synthetic multi-photometry.
This new tool should be publicly available in 2002 on the following 
web site {\tt http://www.astro.mat.uc.pt/BaSeL/}.


\end{document}